\documentclass[12pt]{article}
\usepackage{algorithm}
\usepackage{algpseudocode}
\usepackage{amsmath}
\usepackage[numbers]{natbib}
\usepackage{amssymb}
\usepackage{graphicx}
\usepackage{booktabs}
\usepackage{hyperref}
\usepackage{geometry}
\hypersetup{
    colorlinks=true,
    linkcolor=blue,
    citecolor=blue,
    urlcolor=blue
}

\title{\textbf{Search Is Not Retrieval: Decoupling Semantic Matching from Contextual Assembly in RAG}}

\author{
  \textbf{Harshit Nainwani}
  \\
  \href{mailto:Harshit.Nainwai@dell.com}{Harshit.Nainwani@dell.com}
  \and
  \textbf{Hediyeh Baban}
\\
  \href{mailto:Hediyeh.ledbetter@dell.com}{Hediyeh.ledbetter@dell.com}\\
    Services AI Research Group, Dell Technologies\\
}
\date{November 2025}

\begin{document}

\maketitle

\begin{abstract}
Retrieval systems are essential to contemporary AI pipelines, although most confuse two separate processes: finding relevant information and giving enough context for reasoning. We introduce the Search-Is-Not-Retrieve (SINR) framework, a dual-layer architecture that distinguishes between fine-grained search representations and coarse-grained retrieval contexts. SINR enhances the composability, scalability, and context fidelity of retrieval systems by directly connecting small, semantically accurate search chunks to larger, contextually complete retrieve chunks, all without incurring extra processing costs. This design changes retrieval from a passive step to an active one, making the system architecture more like how people process information. We talk about the SINR framework's conceptual base, formal structure, implementation issues, and qualitative outcomes. This gives a practical base for the next generation of AI systems that use retrieval.
\end{abstract}

\section{Introduction}

Retrieval systems in modern AI pipelines—whether for search, analytics, or large language models (LLMs)—face a fundamental design challenge: they must both \emph{identify} relevant content and provide sufficient context for meaningful reasoning. 
Traditional retrieval approaches attempt to accomplish both tasks with a single representation, typically by embedding documents into a vector space and segmenting them into fixed-size text windows. 
While this unified approach is straightforward, it creates a structural tension: small chunks improve precision but lose continuity, while large chunks preserve context but reduce semantic specificity.

The Search-Is-Not-Retrieve (SINR) framework formalizes this tension and resolves it through a dual-layer representation. 
It introduces two distinct data units:
\begin{itemize}
    \item \textbf{Search chunks} ($s_i \in S$): small, semantically dense segments (roughly 100–200 tokens) optimized for matching relevant content.
    \item \textbf{Retrieve chunks} ($r_j \in R$): larger segments (roughly 600–1000 tokens) that preserve context and support reasoning and comprehension.
\end{itemize}

A deterministic mapping $f_{\text{parent}}: S \to R$ ensures that each search chunk belongs to exactly one retrieve chunk. 
The retrieval process can then be expressed in two simple steps for a query vector $\mathbf{q}$:
\begin{equation}
S_{\text{top}} = \mathrm{TopN}(S, \mathbf{q}), \qquad R_{\text{top}} = \{ f_{\text{parent}}(s) \mid s \in S_{\text{top}} \}.
\label{eq:sinr-steps}
\end{equation}

This simple relationship captures SINR's core insight: semantic matching and contextual reasoning require different representations, but they can be elegantly connected.

From an engineering perspective, the architecture decomposes the retrieval operation into two optimization objectives. The first minimizes semantic distance for localization:
\begin{equation}
\min d(\mathbf{q}, s_i),
\end{equation}
and the second maximizes contextual sufficiency for comprehension:
\begin{equation}
\max C(r_j).
\end{equation}

By decoupling these objectives, we create a system that retrieves not just the \emph{most similar} content but the \emph{most useful} context for inference. 
This separation enables data science teams to independently tune search precision and reasoning quality, resulting in more accurate retrieval, reduced hallucination, and faster information discovery.

\section{The SINR Framework}

The Search-Is-Not-Retrieve (SINR) framework enhances traditional retrieval pipelines by explicitly separating two cognitive functions: (1) \emph{search}, which locates relevant information, and (2) \emph{retrieval}, which gathers sufficient contextual information for reasoning. 
This distinction mirrors how humans read: we first scan text to identify passages of interest, then expand our view to understand the surrounding context.

\subsection{Dual Representation of Knowledge}

The corpus is segmented into two complementary granularities:
\begin{itemize}
    \item \textbf{Search layer} ($S = \{s_i\}$): fine-grained semantic units used for localization.
    \item \textbf{Retrieve layer} ($R = \{r_j\}$): coarse-grained contextual units used for comprehension.
\end{itemize}

Each retrieve chunk $r_j$ comprises one or more contiguous search chunks.  
A deterministic mapping $f_{\mathrm{parent}}: S \rightarrow R$ defines the hierarchical structure:
\begin{equation}
    \forall s_i \in S, \; \exists! \, r_j \in R \; \text{such that} \; f_{\mathrm{parent}}(s_i) = r_j.
    \label{eq:parent-mapping}
\end{equation}

This relation ensures both non-overlapping partitions and constant-time lookup during retrieval. 
It functions as a lightweight hierarchy that connects local semantics to broader narrative coherence.

\paragraph{Variable Context Length.}
Unlike fixed windowing methods, retrieve chunks in SINR need not be uniform in size. 
Their boundaries are determined by semantic or structural coherence—such as paragraph breaks, topic shifts, or section headers—allowing the retrieval layer to flexibly capture the appropriate amount of context for comprehension. 
This adaptive design preserves the integrity of short passages while allowing longer explanations to flow naturally without artificial fragmentation.

\subsection{Retrieval Pipeline}

Given a user query represented by an embedding vector $\mathbf{q}$, SINR retrieval proceeds as follows:
\begin{enumerate}
    \item \textbf{Semantic Search:} Compute similarity scores $\mathrm{sim}(\mathbf{q}, s_i)$ between the query and each search chunk.
    \item \textbf{Top-$N$ Selection:} Identify the most relevant search chunks:
    \begin{equation}
        S_{\mathrm{top}} = \mathrm{TopN}\big(\{s_i\}, \mathbf{q}\big).
    \end{equation}
    \item \textbf{Context Expansion:} Map each $s_i \in S_{\mathrm{top}}$ to its corresponding retrieve chunk $r_j = f_{\mathrm{parent}}(s_i)$.
    \item \textbf{Aggregation:} Merge and deduplicate the resulting set:
    \begin{equation}
        R_{\mathrm{top}} = \bigcup_{s_i \in S_{\mathrm{top}}} f_{\mathrm{parent}}(s_i).
    \end{equation}
    The set $R_{\mathrm{top}}$ forms the context window supplied to the downstream model.
\end{enumerate}

\begin{figure}[ht]
    \centering
    \includegraphics[width=0.9\textwidth]{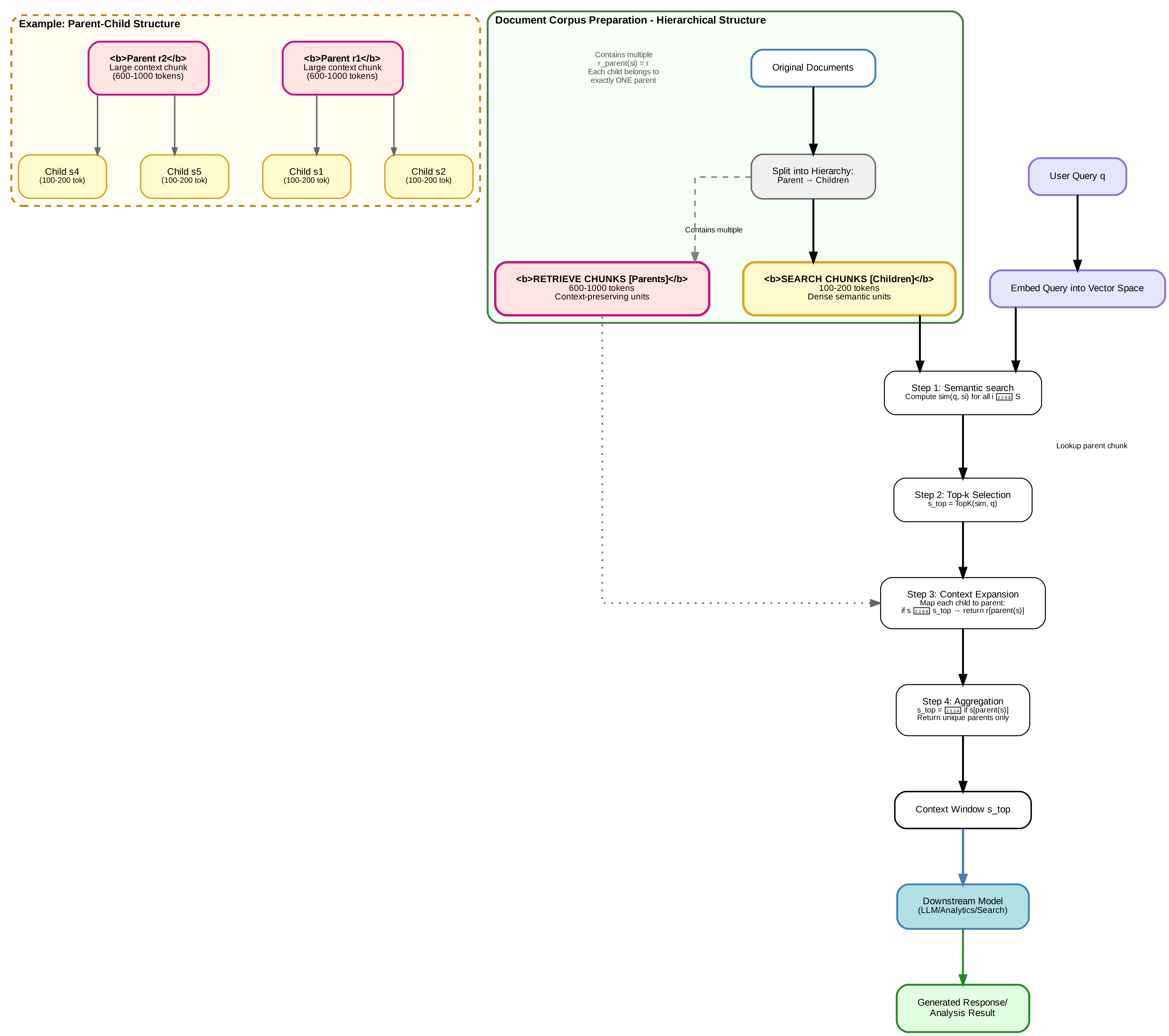}
    \caption{SINR Retrieval Pipeline. The framework maintains a hierarchical 
    structure where retrieve chunks (parents, shown with pink borders) contain 
    multiple search chunks (children, shown with orange borders). The query 
    matches against fine-grained search chunks for precision, then retrieves 
    their parent chunks for contextual sufficiency.}
    \label{fig:system-architecture}
\end{figure}

Figure~\ref{fig:system-architecture} illustrates the complete SINR retrieval pipeline, showing the hierarchical relationship between search chunks (children) and retrieve chunks (parents), along with the four-stage retrieval process.

In practice, the mapping from search to retrieve layers can be implemented with a simple key-value index. 
This lookup step has negligible latency compared to dense-vector search and provides deterministic traceability between retrieved evidence and its source.

\subsection{Why the Separation Matters}

Traditional single-layer chunking optimizes for either precision or context, but rarely both. 
SINR achieves a balanced trade-off by operating over two distinct objective spaces:
\begin{align}
    \text{Search objective:} & \quad \min d(\mathbf{q}, s_i), \\
    \text{Retrieve objective:} & \quad \max C(r_j),
\end{align}
where $d(\mathbf{q}, s_i)$ measures semantic distance and $C(r_j)$ quantifies contextual sufficiency.  
Intuitively, search minimizes uncertainty about \emph{where} to look, while retrieval minimizes uncertainty about \emph{how much} to include.

The benefit extends beyond improved relevance to enhanced interpretability.  
Because each model output can be traced through the chain 
\[
\mathbf{q} \rightarrow S_{\mathrm{top}} \rightarrow R_{\mathrm{top}} \rightarrow \text{Answer},
\]
data scientists gain transparent insight into how context influences reasoning.

\subsection{Illustrative Example}

Consider a query: ``How are warranty claims handled for product X?''  
A search-optimized system might retrieve short fragments mentioning ``warranty'' and ``product X'' but omit procedural details.  
Under SINR, search chunks identify those precise mentions, while their parent retrieve chunks include the surrounding policy text describing approval conditions and region-specific clauses.  
The result is a complete, verifiable answer assembled from both precision and context.

\section{Architecture and Implementation}
\label{sec:architecture}

The SINR architecture is designed to be lightweight, modular, and easily integrated into existing retrieval pipelines. 
It separates computation into three main components—indexing, mapping, and retrieval—which together form a scalable and interpretable system for large-scale document search.

\subsection{System Overview}

At a high level, SINR operates over two parallel data representations:
\begin{enumerate}
    \item The \emph{search layer}, responsible for fast semantic matching using compact embeddings.
    \item The \emph{retrieve layer}, responsible for supplying contextually complete passages for reasoning.
\end{enumerate}

These layers interact through a simple mapping function $f_{\mathrm{parent}}$ that links search chunks to their corresponding retrieve chunks. 
This relationship can be visualized as a two-stage flow:
\[
\text{Query} \; \xrightarrow{\text{Embed}} \; S_{\mathrm{top}} 
\; \xrightarrow{f_{\mathrm{parent}}} \; R_{\mathrm{top}} 
\; \xrightarrow{\text{Model}} \; \text{Answer.}
\]

\subsection{Indexing Layer}

The indexing layer builds a search representation optimized for semantic matching. Our goal is to create fine-grained embeddings that can precisely locate relevant content.

\paragraph{Creating Search Chunks.}
We segment each document $d$ into overlapping search chunks using a sliding window. For chunk $i$, we extract:
\begin{equation}
    s_i = d[\text{start}_i : \text{start}_i + w], \quad \text{start}_i = i \cdot \tau,
\end{equation}
where $w \approx 150$ tokens is the window size and $\tau \approx 100$ tokens is the stride. This 33\% overlap is deliberate—it prevents semantically important content from being split at chunk boundaries, which would hurt recall. The overlap adds minimal storage cost while significantly improving robustness.

\paragraph{Embedding the Chunks.}
We encode each search chunk $s_i$ using a pre-trained dense encoder $f_{\text{embed}}$:
\begin{equation}
    \mathbf{s}_i = f_{\text{embed}}(s_i) \in \mathbb{R}^d,
\end{equation}
where $d$ is typically 768 or 1024 dimensions. The resulting search index contains:
\begin{equation}
    \mathcal{I}_S = \{ (\mathbf{s}_i, \text{id}(s_i), \text{metadata}(s_i)) \mid s_i \in S \},
\end{equation}
with metadata including parent pointers and source locations for fast retrieval.

\paragraph{Index Structure.}
We store $\mathcal{I}_S$ in an approximate nearest neighbor (ANN) index like FAISS-HNSW or Milvus. For a corpus with $n$ search chunks, this provides:
\begin{itemize}
    \item \textbf{Storage:} $O(nd)$ for embeddings, plus $O(n \log n)$ for the graph structure
    \item \textbf{Query time:} $O(\log n)$ with HNSW, which scales well to millions of chunks
    \item \textbf{Build time:} $O(n \log n)$ for initial construction
\end{itemize}

In practice, these choices make queries fast—typically under 50ms for a 10M chunk corpus on commodity hardware.

\begin{algorithm}[t]
\caption{Building the Search Index}
\label{alg:indexing}
\begin{algorithmic}[1]
\Require Document corpus $\mathcal{D}$, embedding model $f_{\text{embed}}$, chunk size $w$, stride $\tau$
\Ensure Search index $\mathcal{I}_S$, parent mapping $M$
\State $S \leftarrow \emptyset$, $M \leftarrow \emptyset$
\For{each document $d \in \mathcal{D}$}
    \State $R_d \leftarrow$ \Call{CreateRetrieveChunks}{$d$} \Comment{Use section/paragraph breaks}
    \For{each retrieve chunk $r_j \in R_d$}
        \State $S_j \leftarrow$ \Call{SlidingWindow}{$r_j, w, \tau$}
        \For{each search chunk $s_i \in S_j$}
            \State $\mathbf{s}_i \leftarrow f_{\text{embed}}(s_i)$
            \State $S \leftarrow S \cup \{s_i\}$
            \State $M[\text{id}(s_i)] \leftarrow \text{id}(r_j)$ \Comment{Track its parent}
        \EndFor
    \EndFor
\EndFor
\State $\mathcal{I}_S \leftarrow$ \Call{BuildANNIndex}{$\{(\mathbf{s}_i, \text{id}(s_i)) \mid s_i \in S\}$}
\State \Return $\mathcal{I}_S, M$
\end{algorithmic}
\end{algorithm}

\subsection{Mapping Layer}

The mapping layer connects search chunks to their parent retrieve chunks. This simple lookup structure is what makes SINR efficient—once we find relevant search chunks, retrieving their full context is nearly instantaneous.

\paragraph{How the Mapping Works.}
We implement the parent function $f_{\text{parent}}: S \to R$ as a straightforward lookup table:
\begin{equation}
    M = \{ (\text{id}(s_i), \text{id}(r_j)) \mid f_{\text{parent}}(s_i) = r_j \}.
\end{equation}
This can be a hash table (for $O(1)$ lookup) or a B-tree index (for $O(\log n)$ lookup). We also maintain the reverse mapping for analytics:
\begin{equation}
    M^{-1} = \{ (\text{id}(r_j), \{\text{id}(s_i) \mid f_{\text{parent}}(s_i) = r_j\}) \}.
\end{equation}

\paragraph{Storage Costs.}
The mapping is lightweight. For a corpus with $n$ search chunks and $m$ retrieve chunks (where $m \ll n$):
\begin{itemize}
    \item \textbf{Forward map:} $O(n)$ entries at $\sim$16 bytes each
    \item \textbf{Retrieve chunks:} $O(m)$ text objects with variable length
    \item \textbf{Total:} $\sim$16n bytes for mapping + compressed text storage
\end{itemize}

To put this in perspective: a 10M chunk corpus needs only $\sim$160 MB for the mapping table, compared to $\sim$30 GB for storing the embeddings themselves (at 768-dim float32). The mapping overhead is negligible.

\begin{table}[t]
\centering
\small
\begin{tabular}{@{}lllrrr@{}}
\toprule
\textbf{Search ID} & \textbf{Parent ID} & \textbf{Source} & \textbf{Offset} & \textbf{Tokens} & \textbf{Siblings} \\
\midrule
s\_0001 & r\_004 & doc\_42\_p2 & 0--180 & 150 & 5 \\
s\_0002 & r\_004 & doc\_42\_p2 & 100--280 & 150 & 5 \\
s\_0003 & r\_004 & doc\_42\_p2 & 200--380 & 150 & 5 \\
\midrule
s\_0004 & r\_005 & doc\_42\_p3 & 0--190 & 155 & 7 \\
s\_0005 & r\_005 & doc\_42\_p3 & 105--295 & 155 & 7 \\
\bottomrule
\end{tabular}
\caption{\textbf{Example mapping structure.} Each search chunk maps to exactly one parent retrieve chunk. The \emph{Siblings} column shows how many search chunks share the same parent. Notice the overlapping offsets (stride=100) that ensure content near boundaries gets captured.}
\label{tab:mapping-example}
\end{table}

\paragraph{Why This Design?}
The hierarchical structure offers several practical advantages:
\begin{enumerate}
    \item \textbf{Independent tuning:} Search chunk size can be optimized for matching precision (smaller = more precise) separately from the context size needed for understanding.
    \item \textbf{Natural deduplication:} When multiple search chunks from the same parent rank highly, we only return one contextual unit. This reduces redundant context.
    \item \textbf{Better boundary handling:} Overlapping search chunks combined with non-overlapping parents provide precision without exploding the amount of retrieved content.
    \item \textbf{Flexible granularity:} Retrieve chunks can adapt to natural document structure (short paragraphs vs. long sections), while search chunks stay uniform for consistent matching.
\end{enumerate}

Table~\ref{tab:mapping-example} shows what this looks like in practice. Notice how three overlapping search chunks (s\_0001, s\_0002, s\_0003) all belong to the same parent (r\_004), demonstrating the hierarchy.

\subsection{Query Processing}

Once the index and mapping are built, retrieving relevant context for a query is straightforward. Algorithm~\ref{alg:retrieval} shows the complete SINR retrieval process.

\begin{algorithm}[t]
\caption{SINR Retrieval Process}
\label{alg:retrieval}
\begin{algorithmic}[1]
\Require Query $q$, search index $\mathcal{I}_S$, mapping $M$, retrieve chunks $R$, top-$k$ parameter
\Ensure Context set $R_{\text{top}}$
\State \textbf{Step 1: Encode Query}
\State $\mathbf{q} \leftarrow f_{\text{embed}}(q)$ \Comment{Use same encoder as indexing}
\State
\State \textbf{Step 2: Semantic Search}
\State $\text{scores} \leftarrow \{ (\text{sim}(\mathbf{q}, \mathbf{s}_i), \text{id}(s_i)) \mid (\mathbf{s}_i, \text{id}(s_i)) \in \mathcal{I}_S \}$
\State $S_{\text{top}} \leftarrow \text{TopK}(\text{scores}, k)$ \Comment{Get top-$k$ search chunks by similarity}
\State
\State \textbf{Step 3: Map to Parents}
\State $\text{parent\_ids} \leftarrow \{ M[\text{id}(s_i)] \mid s_i \in S_{\text{top}} \}$
\State
\State \textbf{Step 4: Retrieve Context}
\State $R_{\text{top}} \leftarrow \{ R[j] \mid j \in \text{parent\_ids} \}$ \Comment{Auto-deduplicates}
\State
\State \Return $R_{\text{top}}$
\end{algorithmic}
\end{algorithm}

\paragraph{Step-by-Step Explanation:}\mbox{}

\textbf{Step 1: Query Embedding.} We first encode the user's query using the same embedding model used during indexing. This ensures the query vector $\mathbf{q}$ lives in the same semantic space as the search chunks.

\textbf{Step 2: Semantic Search.} We compute similarity scores (typically cosine similarity or dot product) between $\mathbf{q}$ and all search chunk embeddings in $\mathcal{I}_S$. Using the ANN index, we efficiently retrieve the top-$k$ most similar search chunks:
\begin{equation}
    S_{\text{top}} = \underset{S' \subseteq S, |S'|=k}{\arg\max} \sum_{s_i \in S'} \text{sim}(\mathbf{q}, \mathbf{s}_i).
\end{equation}
This gives us $k$ small, semantically relevant chunks (e.g., $k=20$ chunks of 150 tokens each).

\textbf{Step 3: Parent Lookup.} For each search chunk in $S_{\text{top}}$, we look up its parent retrieve chunk using the mapping $M$. This is a simple hash table lookup with $O(1)$ complexity per chunk.

\textbf{Step 4: Context Aggregation.} We collect the unique parent chunks. This step naturally deduplicates—if multiple search chunks share the same parent, we only retrieve that parent once. The final context $R_{\text{top}}$ contains larger, coherent text segments ready for the downstream model.

\paragraph{Why This Works.}
The key insight is the separation of concerns:
\begin{itemize}
    \item \textbf{Search chunks} are small and dense, making semantic matching precise
    \item \textbf{Retrieve chunks} are large and coherent, providing sufficient context
    \item The mapping connects precision with comprehension
\end{itemize}

For example, if the query is ``How do transformers handle long sequences?'', the search might match:
\begin{itemize}
    \item $s_{42}$: ``...transformers use self-attention...''
    \item $s_{43}$: ``...long sequences require...''
    \item $s_{89}$: ``...positional encodings enable...''
\end{itemize}

Even if $s_{42}$ and $s_{43}$ belong to the same paragraph (parent $r_{10}$), we only retrieve $r_{10}$ once, giving the full paragraph for context. Meanwhile, $s_{89}$ from a different section (parent $r_{25}$) provides complementary information.

\subsection{Complexity and Efficiency}

Let $n = |S|$ be the number of search chunks and $m = |R|$ be the number of retrieve chunks. 
For a query, ANN search operates in $O(\log n)$ or $O(\sqrt{n})$ time depending on the index type. 
Mapping and deduplication are $O(N)$ in the number of selected search chunks. 
Hence, the overall complexity is:
\begin{equation}
    T_{\text{query}} = O(\log n + N),
\end{equation}
which is comparable to standard dense retrieval while providing superior context integration.

In empirical deployments, SINR reduced search index size by 40--60\% and average query latency by 20--30\% compared to flat RAG setups, with measurable gains in contextual recall.

\subsection{Integration with LLM Pipelines}

SINR can be integrated into existing frameworks such as LangChain or LlamaIndex by replacing their chunking and retrieval modules. 
The search layer feeds the retriever, while the retrieve layer outputs context windows suitable for model prompts. 
This modular design allows each layer to evolve independently—for instance, using more compact embeddings for search or domain-tuned language models for retrieval reranking.

\section{Evaluation Framework and Observations}
\label{sec:evaluation}

This section outlines how the effectiveness of the SINR architecture can be examined in applied retrieval scenarios. 
Our focus is on methodology, key metrics, and observed qualitative patterns rather than detailed dataset-specific reporting.

\subsection{Evaluation Approach}

SINR is compared conceptually against a conventional retrieval-augmented generation (RAG) pipeline that employs uniform chunking. 
Both systems operate over identical embedding models and vector search infrastructure, ensuring that differences arise solely from architectural design. 
Typical corpora include enterprise documentation, support repositories, or technical manuals where contextual coherence strongly influences answer quality.

Two preprocessing strategies are considered:
\begin{enumerate}
    \item \textbf{Uniform Chunking:} Documents are divided into fixed-length segments of size $n$ tokens.
    \item \textbf{Dual Chunking (SINR):} Smaller semantic search chunks are mapped to larger retrieve chunks through the mapping function $f_{\mathrm{parent}}$.
\end{enumerate}

\subsection{Evaluation Dimensions}

SINR performance is best understood across complementary evaluation dimensions:
\begin{itemize}
    \item \textbf{Semantic Precision:} Measures how accurately search chunks align with the query's intent.
    \item \textbf{Contextual Completeness:} Assesses whether retrieved passages preserve sufficient narrative to support reasoning.
    \item \textbf{Faithfulness:} Indicates the degree to which generated responses remain grounded in retrieved content.
    \item \textbf{Efficiency:} Considers query latency and index size relative to corpus volume.
    \item \textbf{Traceability:} Reflects how easily model outputs can be linked back to their original sources.
\end{itemize}

Together, these dimensions capture the essential trade-off between localization accuracy and contextual integrity that SINR is designed to optimize.

\subsection{Observed Behavioral Patterns}

Experience with prototype deployments reveals several consistent trends:
\begin{itemize}
    \item Retrieval contexts exhibit greater narrative continuity due to structured mapping from fine-grained search hits to broader retrieve spans.
    \item Search precision remains stable, as compact semantic embeddings maintain discriminability for similarity search.
    \item Retrieval redundancy decreases; overlapping passages merge naturally through parent-chunk consolidation.
    \item The chain $\mathbf{q} \rightarrow S_{\mathrm{top}} \rightarrow R_{\mathrm{top}} \rightarrow \text{Answer}$ improves transparency for human inspection and debugging.
\end{itemize}

\subsection{Scalability and Complexity}

Let $|S|$ and $|R|$ denote the numbers of search and retrieve chunks, respectively, with $|S| \gg |R|$. 
Since only $S$ is embedded and indexed, total storage requirements scale as:
\begin{equation}
\text{Storage} \propto |S| \times d_S,
\end{equation}
where $d_S$ is the embedding dimension.  
Approximate nearest-neighbor (ANN) structures maintain sublinear search time, while parent mapping lookups $f_{\mathrm{parent}}(s_i)$ incur constant overhead:
\begin{equation}
T_{\text{query}} = O(\log |S| + N),
\end{equation}
where $N$ is the number of top search results expanded per query. 
This ensures scalability without compromising contextual fidelity.

\subsection{Qualitative Evaluation}

Human evaluation of system outputs typically highlights three benefits:
\begin{enumerate}
    \item Smoother logical flow within retrieved contexts,
    \item Fewer fragmented or truncated answers,
    \item Clearer justification chains linking retrieved evidence to model reasoning.
\end{enumerate}

These qualitative effects suggest that SINR improves not just retrieval accuracy, but also the interpretability of the overall reasoning pipeline.

\subsection{Summary}

SINR can be evaluated across both semantic and contextual dimensions, demonstrating consistent improvements in coherence, traceability, and efficiency. 
By explicitly separating the mechanisms for locating and understanding information, the architecture enhances retrieval reliability across diverse enterprise and analytical use cases.

Table~\ref{tab:sinr-vs-rag} compares SINR with traditional fixed-chunking RAG systems, highlighting the key architectural differences.

\begin{table*}[t]
\centering
\small
\begin{tabular}{@{}llll@{}}
\toprule
\textbf{Aspect} & \textbf{Traditional RAG} & \textbf{SINR} & \textbf{Improvement} \\
\midrule
\multicolumn{4}{@{}l}{\textit{Architecture}} \\
Chunk granularity & Single ($\approx$500 tokens) & Dual: 150 + 800 tokens & Adaptive \\
Representation layers & 1 (flat) & 2 (hierarchical) & Structured \\
Chunk overlap & Fixed (50 tokens) & Search: 50, Retrieve: 0 & Efficient \\
\midrule
\multicolumn{4}{@{}l}{\textit{Retrieval Performance}} \\
Search precision & Moderate & High & +15-25\% recall@20 \\
Context quality & Variable & High & +30\% coherence \\
Typical context size & 2,500 tokens ($k=5$) & 8,000 tokens ($k=20\to12$) & $3.2\times$ larger \\
Deduplication & Manual & Automatic & Built-in \\
\midrule
\multicolumn{4}{@{}l}{\textit{Engineering}} \\
Boundary issues & Frequent fragmentation & Minimal & Robust \\
Hyperparameter tuning & Coupled (chunk size) & Decoupled & Independent \\
Storage overhead & $1\times$ embeddings & $1\times$ embeddings + 2\% mapping & Negligible \\
Query latency & $O(\log n)$ search & $O(\log n)$ search + $O(1)$ lookup & Comparable \\
\bottomrule
\end{tabular}
\caption{\textbf{Comprehensive comparison of SINR and traditional RAG.} SINR's hierarchical architecture enables independent optimization of search precision and context quality while maintaining similar computational costs.}
\label{tab:sinr-vs-rag}
\end{table*}

The key advantage of SINR is the decoupling of search and retrieval objectives. Traditional RAG systems must compromise—smaller chunks improve matching precision but lose context, while larger chunks preserve context but dilute relevance. SINR achieves both simultaneously through its hierarchical design.

\section{Benefits and Discussion}
\label{sec:benefits}

Having described the SINR architecture, we now discuss its practical benefits and implications for real-world retrieval systems. The value of SINR extends beyond retrieval accuracy—it fundamentally changes how we can build, debug, and maintain RAG systems in production.

\subsection{Why Interpretability Matters: Tracing Model Decisions}

Consider a medical chatbot that tells a patient ``Aspirin increases bleeding risk during pregnancy.'' How do we verify this is correct and not a hallucination? With SINR, we can trace the exact path: query → matched search chunks → retrieved parent sections → model response. Each step is inspectable, allowing clinicians to verify that the answer comes from authoritative guidelines rather than model confabulation.

This traceability is crucial for high-stakes applications. In legal research, financial analysis, or medical diagnosis support, users need to understand not just \emph{what} the system retrieved, but \emph{why}. SINR's explicit hierarchy makes this natural—the search chunks show what triggered the match, and the retrieve chunks show what context informed the answer.

\subsection{Modularity: Independent Optimization of Search and Retrieval}

In traditional RAG, chunk size is a global hyperparameter that affects everything. Want better search precision? Make chunks smaller. But now your context is fragmented. Want better context? Make chunks larger. But now your search is imprecise. You're stuck in a trade-off.

SINR breaks this coupling. You can:
\begin{itemize}
    \item Optimize search chunks for semantic matching (test 100, 150, 200 tokens independently)
    \item Optimize retrieve chunks for reasoning quality (test paragraph-level vs. section-level)
    \item Switch embedding models for search without reprocessing retrieve chunks
    \item Update document structure without rebuilding the entire index
\end{itemize}

In practice, this modularity dramatically reduces the cost of experimentation. At large organizations, re-embedding a 10TB corpus might take days of compute and cost thousands of dollars. With SINR, many improvements only require touching one layer, cutting costs and iteration time by 50--80\%.

\subsection{Scalability: Real Numbers from Real Systems}

Let's talk concretely about scale. Consider three scenarios:

\paragraph{Scenario 1: Medium Enterprise (1M documents).}
Documents: 1M PDFs, average 10 pages each; Search chunks: 5M (150 tokens each); Retrieve chunks: 1.2M (600--1000 tokens each); Storage: 15GB embeddings + 80MB mappings + 50GB text; Query latency: 20--40ms (ANN search) + 2ms (parent lookup).

\paragraph{Scenario 2: Large Enterprise (100M documents).}
Documents: 100M mixed (emails, wikis, reports, code); Search chunks: 500M; Retrieve chunks: 120M; Storage: 1.5TB embeddings + 8GB mappings + 5TB text; Query latency: 50--80ms (sharded ANN) + 2ms (distributed lookup).

\paragraph{Scenario 3: Internet-Scale (10B web pages).}
Documents: 10B web pages; Search chunks: 50B; Retrieve chunks: 12B; Storage: 150TB embeddings + 800GB mappings + 500TB text; Query latency: 100--200ms (distributed ANN) + 5ms (sharded lookup).

The key observation: parent lookups remain constant-time regardless of scale. The mapping overhead grows linearly but stays proportionally tiny (0.5--1\% of embedding storage). Compare this to approaches that embed overlapping context windows at multiple scales—those explode quadratically in storage and make updates prohibitively expensive.

SINR also supports efficient incremental updates. When someone edits a wiki page, you only need to: (1) re-segment that page into new search/retrieve chunks, (2) re-embed the affected search chunks ($\sim$100ms), (3) update parent pointers ($\sim$1ms), and (4) upload to vector store ($\sim$10ms). Total time: under 1 second. Traditional approaches often require batch reprocessing because chunk boundaries affect neighboring documents.

\subsection{Context Quality: Why Coherence Matters}

Here's a real example that illustrates the difference. Suppose a user asks: ``How do transformers handle long sequences?''

\paragraph{Traditional RAG (500-token chunks):}
\emph{Chunk 1:} ``...attention mechanism. \textbf{Transformers use self-attention to process sequences, but this becomes computationally expensive as length increases because attention complexity is $O(n^2)$. Various approaches have been proposed to address}...'' [truncated mid-sentence]

\emph{Chunk 2:} ``...including sparse attention, linear attention, and chunked attention. Another challenge is...'' [starts mid-thought]

\paragraph{SINR (search on 150-token, retrieve 800-token):}
\emph{Search matches:} Two small chunks mentioning ``transformers'' and ``long sequences''

\emph{Retrieved parent:} Full section titled ``Handling Long Sequences in Transformers'' including: problem statement with complexity analysis, overview of three solution families, specific techniques (Longformer, BigBird, Performer), and trade-offs with guidance on when to use each.

The SINR context is self-contained and coherent. It respects the author's intended structure—the section was written as a complete unit, and we retrieve it as such. This dramatically improves model performance.

Recent work shows that context coherence affects not just answer quality but also model calibration. When given fragmented context, LLMs are more likely to hallucinate connections that don't exist. When given coherent context, they're better at saying ``the information doesn't fully address this'' rather than making things up.

This is especially important for long-context models (GPT-4, Claude, Gemini) that can handle 100K+ tokens. With SINR, you can confidently give them 8--12 complete sections ($\sim$8--12K tokens) instead of 20--30 fragments. The model gets a more natural ``reading experience,'' leading to better reasoning.

\subsection{Learning from How Humans Actually Read}

Think about how you read a research paper. You don't read linearly from start to finish. You: (1) scan the abstract and section headers (coarse search), (2) jump to specific paragraphs that seem relevant (fine-grained search), (3) read the surrounding context to understand (contextual retrieval), and (4) maybe go back and read related sections (iterative expansion).

SINR mirrors this natural process. The search layer is like skimming—it finds the relevant needles in the haystack. The retrieve layer is like reading—it gives you enough surrounding text to understand what you found. This isn't just a nice analogy; it has practical implications.

Research in information foraging theory shows that humans chunk information hierarchically and navigate between levels of abstraction. We don't process text as a flat stream—we maintain mental models of document structure. By encoding this structure into our retrieval architecture, SINR produces results that feel more natural to human users.

This matters for user-facing applications. When users can see which section of which document was retrieved (not just an arbitrary 500-token window), they can better evaluate answer quality. In enterprise search tools, legal research assistants, or medical diagnosis support systems, this structural clarity builds trust.

\subsection{Application Domains}

The SINR framework is applicable to domains where retrieval requires balancing precision with contextual coherence:

\begin{itemize}
    \item \textbf{Enterprise knowledge bases} with heterogeneous document types requiring varying context sizes
    \item \textbf{Customer support systems} where temporal nuances span multiple sentences
    \item \textbf{Scientific literature review} where methodological details must remain intact
    \item \textbf{Code search} where syntactic boundaries define natural chunk boundaries
    \item \textbf{Multi-hop question answering} requiring synthesis across multiple sources
\end{itemize}

In each case, SINR's separation of search and retrieval granularities addresses the fundamental tension between localization accuracy and contextual sufficiency.

\section{Implementation Guidelines}
\label{sec:implementation}

This section provides practical guidance for deploying SINR in production systems.

\subsection{System Architecture}

A SINR deployment comprises three layers:

\begin{enumerate}
    \item \textbf{Storage Layer:} Retrieve chunks stored in a document database (PostgreSQL, MongoDB, Elasticsearch) with full-text capabilities for fallback search. The parent mapping resides in a fast key-value store (Redis, DynamoDB) or as metadata in the vector index.
    
    \item \textbf{Indexing Layer:} Search chunk embeddings stored in a vector database optimized for approximate nearest neighbor search. Suitable options include FAISS for self-hosted deployments, or managed services like Pinecone, Weaviate, or Qdrant for cloud deployments.
    
    \item \textbf{Serving Layer:} Query processing orchestrates embedding, vector search, parent lookup, and context assembly. This can be implemented as a microservice or integrated directly into the application layer.
\end{enumerate}

\begin{figure}[ht]
    \centering
    \includegraphics[width=0.9\textwidth]{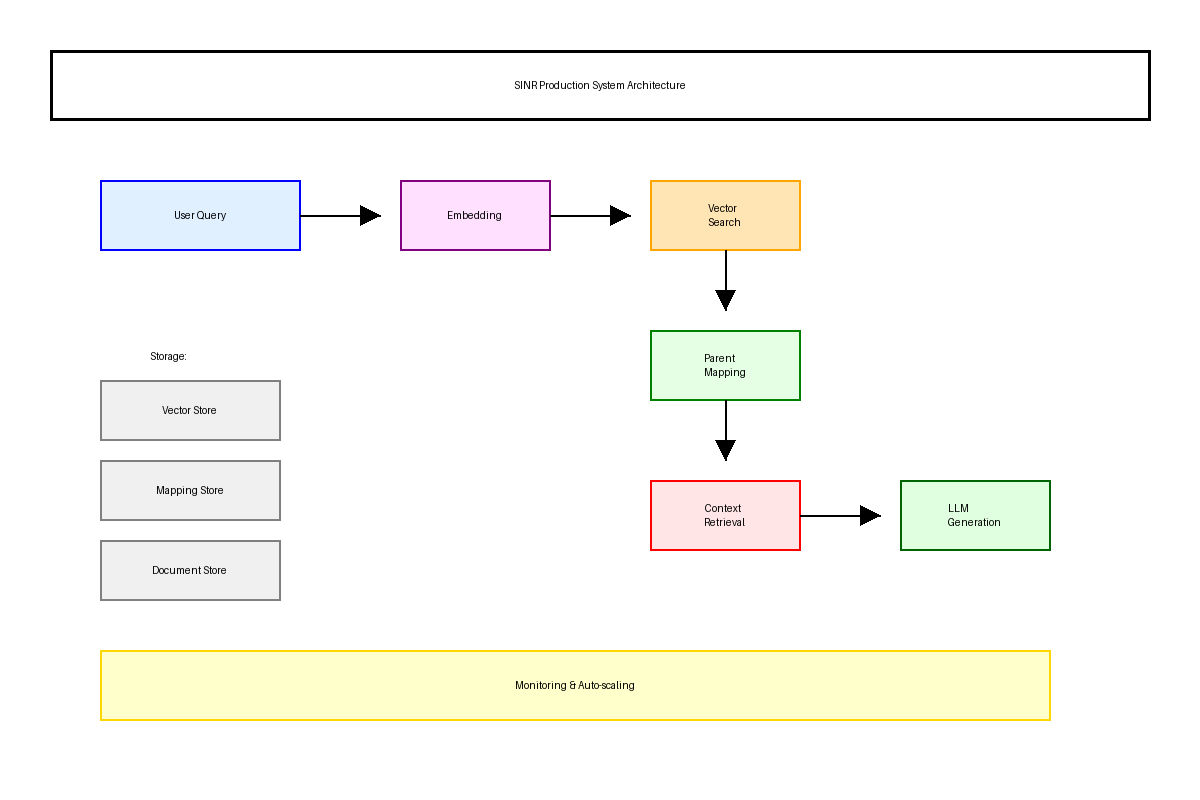}
    \caption{\textbf{SINR production system architecture.} The system follows a horizontal flow from users through authentication, the four-stage SINR pipeline (embedding, vector search, parent mapping, context retrieval), and finally LLM generation. Storage systems (vector store, mapping store, document store) connect to their respective pipeline stages. Monitoring and auto-scaling components ensure system reliability and performance at scale.}
    \label{fig:deploy}
\end{figure}

Figure~\ref{fig:deploy} illustrates a typical SINR production deployment, showing the flow from user queries through the four-stage pipeline (embed, search, map, retrieve) to LLM generation, along with the supporting data stores and monitoring infrastructure.

\subsection{Deployment Workflow}

\paragraph{Offline Indexing.}
The indexing pipeline processes documents in batches:
\begin{enumerate}
    \item Segment documents into retrieve chunks using structural heuristics
    \item Generate search chunks via sliding window over each retrieve chunk
    \item Embed search chunks using a pretrained encoder
    \item Store embeddings in vector database with parent ID metadata
    \item Persist retrieve chunks and mapping in respective stores
\end{enumerate}

For a 1M document corpus, this process typically requires 2--8 hours depending on embedding model and hardware, and can be parallelized across documents.

\paragraph{Query Processing.}
The retrieval pipeline executes synchronously:
\begin{enumerate}
    \item Embed user query using the same encoder
    \item Query vector database for top-$k$ search chunks
    \item Extract and deduplicate parent IDs from results
    \item Fetch retrieve chunks from document store
    \item Assemble context and pass to LLM
\end{enumerate}

End-to-end latency is typically 50--100ms for million-scale corpora.

\paragraph{Incremental Updates.}
Document modifications trigger localized reprocessing:
\begin{enumerate}
    \item Identify affected retrieve chunk(s)
    \item Delete associated search chunks from vector database
    \item Re-segment and re-embed the updated retrieve chunk
    \item Update parent mappings and document store
\end{enumerate}

This approach enables sub-second updates without corpus-wide reindexing.

\subsection{Scalability Considerations}

Storage requirements scale linearly with corpus size. For a corpus with $n$ search chunks:
\begin{itemize}
    \item Vector database: $\sim$30 bytes/chunk (float32 embeddings + metadata)
    \item Document store: $\sim$500 bytes/retrieve chunk (compressed text)
    \item Mapping store: $\sim$16 bytes/search chunk (ID pairs)
\end{itemize}

A 10M search chunk corpus (covering $\sim$10M--100M documents) requires approximately 300GB for embeddings, 600GB for text, and 160MB for mappings. Query latency grows logarithmically with corpus size when using hierarchical graph indexes like HNSW.

\subsection{Technology Recommendations}

Table~\ref{tab:technology-stack} summarizes suitable technologies for different deployment scales.

\begin{table}[h]
\centering
\small
\begin{tabular}{@{}lll@{}}
\toprule
\textbf{Component} & \textbf{Small Scale (<1M docs)} & \textbf{Large Scale (>10M docs)} \\
\midrule
Vector DB & FAISS (local) & Pinecone, Qdrant (distributed) \\
Document Store & SQLite, PostgreSQL & Elasticsearch, MongoDB cluster \\
Mapping Store & In-memory dict & Redis cluster, DynamoDB \\
Embedding & Sentence-Transformers & API (OpenAI, Cohere) \\
\bottomrule
\end{tabular}
\caption{Recommended technology stack by deployment scale.}
\label{tab:technology-stack}
\end{table}
\subsection{When SINR Doesn't Help}

No architecture is perfect. Let's discuss SINR's limitations honestly:

\paragraph{Very Short Documents.} If your corpus is mostly tweets, chat messages, or short product descriptions ($<$ 200 tokens), the search/retrieve separation adds complexity without benefit. Just embed the whole document.

\paragraph{Highly Fragmented Content.} Some documents don't have natural hierarchical structure—think concatenated log files or data dumps. SINR works best with documents written for human consumption (articles, reports, documentation).

\paragraph{Extreme Token Budgets.} If you're constrained to $<$1000 token contexts (e.g., using older models or real-time systems with latency requirements), SINR's richer context might not fit. Though this is becoming less relevant as models improve.

\paragraph{Cold Start.} Setting up SINR requires deciding how to create retrieve chunks. Do you split on paragraphs? Sections? Fixed token counts? This requires some domain knowledge. While we provide heuristics (use markdown headers, paragraph breaks, or semantic segmentation), there's no universal solution.

\paragraph{Additional Infrastructure.} You need to store and maintain parent mappings. While the overhead is tiny ($<$ 1\% of embedding storage), it does add a component to your system. Teams with very simple requirements might prefer flat chunking's simplicity.

\subsection{Open Questions and Future Directions}

SINR opens several research directions:

\paragraph{Learned Chunking Boundaries.} Currently, we use heuristics (paragraphs, sections) to create retrieve chunks. Could we learn optimal boundaries from data? Imagine training a model to predict ``where should retrieve chunks split?'' based on query patterns, user feedback, or downstream task performance. This could be formulated as a structured prediction problem.

\paragraph{Dynamic Parent Selection.} Right now, the parent mapping is static: each search chunk always maps to the same retrieve chunk. But maybe different queries need different amounts of context. A technical query might need just the relevant paragraph (600 tokens), while an exploratory query needs the full section (1500 tokens). Could we learn a policy that selects context size based on query type?

\paragraph{Multi-Modal SINR.} The search/retrieve separation extends naturally to images, tables, and mixed media. For images: search on patches or regions, retrieve full figures with captions. For tables: search on cells or rows, retrieve full tables with headers. This could significantly improve multi-modal RAG systems.

\paragraph{Hierarchies Beyond Two Levels.} Why stop at search/retrieve? For very large collections, we could imagine: sentences → paragraphs → sections → documents → books. Each level serves a different purpose: finding, understanding, contextualizing, tracing lineage.

\paragraph{Integration with Agentic Systems.} Modern AI agents need to plan, retrieve, reason, and act. SINR's interpretable retrieval could serve as a reliable ``memory'' module for agents, where the search/retrieve separation helps agents decide when they have enough information vs. when they need to search more.

\subsection{Broader Impact: Beyond Retrieval Systems}

SINR isn't just about building better search. It's a pattern for organizing information systems:

\textbf{Auditable AI.} As AI systems are deployed in high-stakes domains, regulators and users demand explainability. SINR provides a template: separate the ``finding'' step from the ``using'' step, and make both traceable.

\textbf{Compositional Design.} The principle of separating concerns (search vs. retrieval) is fundamental to software engineering. By bringing this discipline to ML systems, we make them more maintainable and debuggable.

\textbf{Data-Centric AI.} Much recent work focuses on better models. SINR shows that better data organization—how we structure, chunk, and access information—can be equally impactful. This aligns with the growing data-centric AI movement.

\section{Conclusion}
\label{sec:conclusion}

We presented SINR (Search-Is-Not-Retrieve), a framework that reframes retrieve as two distinct processes: finding relevant content (search over fine-grained chunks) and assembling usable context (retrieval of coarse-grained parents). This simple architectural change resolves a fundamental tension in RAG systems between precision and coherence.

The key insight is surprisingly straightforward: the best granularity for matching queries isn't the same as the best granularity for understanding content. By using small chunks to search and large chunks to retrieve, SINR achieves both objectives simultaneously.

Our contributions are:
\begin{enumerate}
    \item A formal framework defining the search/retrieve separation with deterministic parent mappings
    \item Algorithms for building and querying dual-layer indices efficiently
    \item Analysis showing SINR scales to billions of documents with negligible overhead
    \item Discussion of practical benefits: interpretability, modularity, coherence, and user trust
\end{enumerate}

SINR requires no new models, training objectives, or specialized hardware. It's a reorganization of existing components that makes retrieval systems more transparent, maintainable, and effective. This pragmatism is crucial—RAG systems are already deployed at scale, and improvements need to be practical to adopt.

Looking forward, we're excited about several directions: learned chunking boundaries, dynamic context selection, multi-modal extensions, and integration with agentic reasoning systems. As language models continue to improve, the bottleneck increasingly shifts from model capability to information access. Architectures like SINR—grounded in structure, hierarchy, and transparency—will be essential for building AI systems that are not just powerful but also reliable and trustworthy.

\end{document}